\documentclass[aps,prl,twocolumn,nobibnotes,showpacs,superscriptaddress]{revtex4-1} 
\usepackage{graphicx} 
\usepackage{dcolumn} 
\usepackage{bm} 

\usepackage{amssymb} 
\usepackage{amsmath} 
\usepackage[alsoload=hep]{siunitx} 
\usepackage{color}
\usepackage{xcolor}
\usepackage{hyperref}

\newcommand{\ee}{\ifmmode (e,e') \else $(e,e')$~\fi}
\newcommand{\Aee}{\ifmmode A(e,e') \else $A(e,e')$~\fi}
\newcommand{\eep}{\ifmmode (e,e'p) \else $(e,e'p)$~\fi}
\newcommand{\Aeep}{\ifmmode A(e,e'p) \else $A(e,e'p)$~\fi}
\newcommand{\pmiss}{\ifmmode p_{miss} \else $p_{miss}$\fi}
\newcommand{\emiss}{\ifmmode E_{miss} \else $E_{miss}$\fi}

\newcommand{\het}{\ifmmode ^3{\rm He} \else $^3$He\fi}
\newcommand{\trit}{\ifmmode ^3{\rm H} \else $^3$H\fi}
\newcommand{\hets}{\ifmmode ^3{\rm He} \else $^3$He \fi} 
\newcommand{\trits}{\ifmmode ^3{\rm H} \else $^3$H  \fi} 
\begin{document}

\title{Probing few-body nuclear dynamics via $^3$H and $^3$He \bm{$\eep$}pn cross-section measurements}

\author{R. Cruz-Torres} \thanks{Equal Contribution} \affiliation{Massachusetts Institute of Technology, Cambridge, MA}
\author{D.~Nguyen} \thanks{Equal Contribution} \affiliation{Massachusetts Institute of Technology, Cambridge, MA} \affiliation{University of Education, Hue University, Hue City, Vietnam} 
\author{F. Hauenstein} \affiliation{Old Dominion University, Norfolk, VA}
\author{A. Schmidt} \affiliation{Massachusetts Institute of Technology, Cambridge, MA}
\author{S. Li} \affiliation{University of New Hampshire, Durham, NH}
\author{D.~Abrams} \affiliation{University of Virginia, Charlottesville, VA}
\author{H.~Albataineh} \affiliation{Texas A \& M University, Kingsville, TX}
\author{S.~Alsalmi } \affiliation{King Saud University, Riyadh, Kingdom of Saudi  Arabia}
\author{D.~Androic} \affiliation{University of Zagreb, Zagreb, Croatia}
\author{K.~Aniol} \affiliation{California State University , Los Angeles, CA}
% ...

\collaboration{Jefferson Lab Hall A Tritium Collaboration}

\date{today}

\begin{abstract}
  We report the first measurement of the \eep three-body breakup
  reaction cross sections in helium-3 (\het) and tritium (\trit) at
  large momentum transfer ($\langle Q^2 \rangle \approx 1.9$
  (GeV/c)$^2$) and $x_B>1$ kinematics, where the cross section should
  be sensitive to quasielastic (QE) scattering from single nucleons.
  The data cover missing momenta $40 \le \pmiss \le \SI{500}{\mega\eVperc}$
  that, in the QE limit with no rescattering, equals the initial momentum of the probed nucleon.
  The measured cross sections are compared with state-of-the-art
  ab-initio calculations.
Overall good agreement, within
  $\pm20\%$, is observed between data and
  calculations for the full \pmiss{} range for \trit{} and for
  $100 \le \pmiss \le \SI{350}{\mega\eVperc}$ for \het.  Including the
  effects of rescattering of the outgoing nucleon improves agreement
  with the data at $\pmiss > \SI{250}{\mega\eVperc}$ and suggests
  contributions from charge-exchange (SCX) rescattering.
The isoscalar sum of
  \het{} plus \trit, which is largely insensitive to 
  SCX, is described by calculations to within the accuracy of
  the data over the entire \pmiss{}
  range.    
  This validates current models of the ground state of the three-nucleon
system up to very high initial nucleon momenta of $\SI{500}{\mega\eVperc}$.
\end{abstract}

\maketitle



Understanding the structure and properties of nuclear systems is a
formidable challenge with implications ranging from the formation of
elements in the universe to their application in laboratory
measurements of fundamental interactions.  Due to the complexity of
the strong nuclear interaction, nuclear systems are often described
using effective models that are based on various levels of
approximations.  Testing and benchmarking such approximations is a
high priority of modern nuclear physics research.

Measurements of high-energy quasi-elastic (QE) electron scattering
serve a unique role in this endeavor as they can be particularly
sensitive to ground state properties of
nuclei~\cite{Kelly:1996hd}. However, in many studies this sensitivity
is reduced by the lack of exact nuclear ground-state calculations and
by the contribution of non-QE processes to the measured
cross-sections.  Calculations of non-QE contributions are highly
model-dependent and can change the resulting cross-sections
dramatically, hindering the interpretation of measurements in terms of
the nuclear ground state~\cite{Ford:2014yua}.

Studies of the three nucleon system can avoid these issues as 
(A) their ground states are exactly calculable from nuclear-interaction models, and (B) proper choice 
of kinematics can suppress cross-section contributions from
non-QE processes, allowing one to to directly relate 
measured cross-sections to the ground-state momentum distribution.
Thus electron scattering studies of helium-3 (\het) and tritium (\trit) nuclei
can serve as a precision test of modern nuclear theory~\cite{Golak:2005iy}.

While vast amounts of modern electron scattering data on \het{} exist~\cite{Sick:2001rh, Benmokhtar:2004fs, Rvachev:2004yr, Long:2019iig, Mihovilovic:2014gdi, Mihovilovic:2018fux, Zhang:2015kna, Camsonne:2016ged, Riordan:2010id}, 
\trit{} data are very sparse due to the safety limitations associated with 
placing a radioactive tritium target in a high-current electron beam.
Current world data dates back to the early 60's~\cite{Collard:1963zza, Schiff:1963zzb, PhysRev.136.B1030, RevModPhys.37.402} and late 80's~\cite{Dow:1986auc, Beck:1986rhj, Beck:1987zz, Dow:1988rk, Juster:1985sd, Amroun:1994qj}.

This letter reports the first electron scattering cross sections on
\trit{} to be published in over 30 years.  Specifically, we study the
distributions of protons in \het{} and in \trit{} using measurements
of high-energy QE proton knockout reactions in comparison with
predictions of state-of-the-art ab-initio calculations to test their modeling
of the three-nucleon ground state up to very large initial momenta.  
The simultaneous measurement of both \het{} and \trit\eep cross
sections, at the kinematics of our experiment, places
stringent constraints on the possible contribution of non-QE reaction
mechanisms to our measurement, thereby improving the equivalence between
the measured missing momenta and initial nucleon momenta and
increasing its sensitivity to the properties of the \het{} and \trit{} ground-states.

Our measured cross sections are well described by theoretical
calculations to about $20\%$, without the need to include non-QE
processes. This is a great improvement over recent \het\eep
measurements~\cite{Benmokhtar:2004fs} that were dominated by non-QE
processes and were therefore significantly less sensitive to its
ground state, especially at large missing momentum.  Our \trit{} data
is better described by calculations than \het. Including leading
nucleon rescattering improves the agreement between the calculations
and the data. The remaining small difference between data and theory
has the opposite trend for \het{} and \trit{}, which could suggest
residual contributions from single charge exchange (SCX)
processes. The effects of SCX are largely suppressed in the isoscalar
sum of \het{} + \trit{} cross sections, which is described by
calculations to within the accuracy of our data. We thus confirm
modeling of the three-nucleon system up to very high nucleon
momenta of $500$~MeV$/c$.

The experiment ran in 2018 in Hall A of the Thomas Jefferson National
Accelerator Facility (JLab).  It used the two high-resolution
spectrometers (HRSs)~\cite{Alcorn:2004sb} and a 20 $\mu A$ electron
beam at 4.326 GeV incident on one of four identical 25-cm long gas
target cells filled with hydrogen ($70.8 \pm 0.4$ mg/cm$^2$),
deuterium ($142.2 \pm 0.8$ mg/cm$^2$), helium-3 ($53.4 \pm 0.6$
mg/cm$^2$), and tritium ($85.1 \pm 0.8$
mg/cm$^2$)~\cite{Santiesteban:2018qwi}.  
To minimize systematic uncertainties between measurements, the HRS
were not moved when changing among the targets, which were installed
on a linear motion target ladder.

Each HRS consisted of three quadrupole magnets for focusing and one
dipole magnet for momentum analysis, followed by a detector package
consisting of a pair of vertical drift chambers used for tracking and two
scintillation counter planes that provided timing and trigger
signals. A CO$_{2}$ Cherenkov detector placed between the
scintillators and a lead-glass calorimeter placed after them were used
for particle identification.
This configuration is slightly updated with respect to the one in Ref.~\cite{Alcorn:2004sb}.

Scattered electrons were detected in the left-HRS, positioned at
central momentum and angle of $\vert\vec p_e {\!'}\vert  = 3.543$~GeV$/c$ and
$\theta_e = 20.88^{\circ}$, giving a central four-momentum transfer
$Q^2 =\vec q\thinspace^2 -\omega^2 = 2.0$ (GeV$/c$)$^2$ (where the
momentum transfer is $\vec q = \vec p_e - \vec p_e {\!'}$), energy
transfer $\omega = E_{beam} - \vert\vec p_e {\!'}\vert = 0.78$ GeV, and $x_{B} \equiv \frac{Q^2}{2m_p\omega}
= 1.4$ (where $m_p$ is the proton mass).  Knocked-out protons were
detected in the right-HRS at two central settings of
$(\theta_p, p_p)$ = ($48.82^{\circ}$, 1.481~GeV$/c$), and
($58.50^{\circ}$, 1.246~GeV$/c$) corresponding to low-\pmiss\ ($40 \le
\pmiss \le 250$~MeV$/c$) and high-\pmiss\ ($250 \le \pmiss \le 500$
MeV/c), respectively, where $\vec p_{miss} = {\vec p}_p - \vec q$.  
The exact electron kinematics for each \pmiss\
bin varied within the spectrometer acceptances.

In the Plane-Wave Impulse Approximation (PWIA) for QE scattering,
where a single exchanged photon is absorbed on a single proton and the
knocked-out proton does not re-interact as it leaves the nucleus, the
cross section is proportional to the spectral function, the
probability of finding a proton in the nucleus with initial momentum
$\vec p_i$ and separation energy $E_i$.  The momentum distribution is then
the integral of the spectral function over $E_i$: $n(p_i)=\int S(p_i,E_i)dE_i$.  In PWIA,
the
missing momentum and energy equal the initial momentum and separation
energy of the knocked-out nucleon: $\vec{p}_i = \vec{p}_{miss}$, $E_i
= E_{miss}$, where $E_{miss} = \omega - T_p - T_{A-1}$, $T_{A-1} =
(\omega + m_A - E_p) - \sqrt{(\omega + m_A -
  E_p)^2-|\vec{p}_{miss}|^2}$ is the reconstructed kinetic energy of
the residual $A-1$ system. $T_p$ and $E_p$ are the measured
kinetic and total energies of the outgoing proton.

Non-QE reaction mechanisms that lead to the same measured final state
also contribute to the cross section, complicating this simple
picture.  Such mechanisms include rescattering of the struck nucleon
(final-state interactions or FSI), meson-exchange currents (MEC),
and exciting isobar configurations (IC).  In addition, relativistic
effects can be significant
\cite{gao00,udias99,AlvarezRodriguez:2010nb}.

The kinematics of our measurement were chosen to reduce contributions
from such non-QE reaction mechanisms.  For high-$Q^2$ reactions, the
effects of FSI were shown to be reduced by choosing kinematics where
the angle between $\vec{p}_{recoil}=-\vec{p}_{miss}$ and $\vec{q}$ is
$\theta_{rq} \lesssim 40^{\circ}$, which also corresponds to $x_B \ge
1$
~\cite{Boeglin:2011mt,Sargsian:2001ax,Frankfurt:1996xx,Jeschonnek:2008zg,Laget:2004sm,Sargsian:2009hf,Hen:2014gna}.
Additionally MEC and IC were shown to be suppressed for $Q^2 > 1.5$
(GeV/$c$)$^2$ and $x_B > 1$~\cite{Sargsian:2001ax,Sargsian:2002wc}.


The  data analysis follows that previously reported in
Ref.~\cite{Cruz-Torres:2019bqw} for the \het/\trit$(e,e'p)$
cross-section ratio extraction.
We selected electrons by requiring that the particle deposits more
than half of its energy in the calorimeter: $E_{cal} / |\vec{p}|
> 0.5$. We selected \eep coincidence events by placing 
$\pm3\sigma$ cuts around the relative electron and proton event times
and the relative electron and proton  reconstructed target vertices
(corresponding to a $\pm 1.2$ cm cut).
Due to the low experimental luminosity, the random coincidence event
rate was negligible. We discarded a small number of runs with
anomalous event rates.

Measured electrons were required to originate within the central $\pm 9$~cm
of the gas target to exclude events originating from the target walls.
By measuring scattering from an empty-cell-like target we determined that the 
target cell wall contribution to the measured \eep event yield was negligible ($\ll 1\%$).

To avoid the acceptance edges of the spectrometer, we only
analyzed  events that were detected within $\pm 4\%$ of the central
spectrometer momentum, and $\pm \SI{27.5}{\milli\radian}$ in in-plane
angle and $\pm \SI{55.0}{\milli\radian}$ in out-of-plane angle
relative to the center of the spectrometer acceptance. 
We further required
$\theta_{rq} < 37.5^{\circ}$ to minimize the effect of FSI and, in the
high-\pmiss\ kinematics, $x_B > 1.3$ to further suppress non-QE
events.

The spectrometers were calibrated using sieve slit measurements to
define scattering angles and by measuring the kinematically
over-constrained exclusive $^1$H\eep and $^2$H$(e,e'p)n$ reactions.  The
$^1$H\eep reaction \pmiss{} resolution was better than 9 MeV/$c$.  We
verified the absolute luminosity normalization by comparing the
measured elastic $^1$H\ee yield to a parametrization of the world data
\cite{Lomon:2006xb}.  We also found excellent agreement between the
elastic $^1$H\eep and $^1$H\ee rates, confirming that the coincidence trigger
performed efficiently.

One significant difference between \het\eep and \trit\eep stems
from their possible final states. The \trit\eep reaction can only result 
 in a three-body $pnn$ continuum state, while \het{} can break up into either a
two-body $pd$ state or a three-body $ppn$ continuum state.  To allow for a
more detailed comparison of the two nuclei we only considered
three-body breakup reactions by requiring $\emiss > 8$ MeV (i.e.,
above the \het{} two-body breakup peak). 


The cross section was calculated from the \eep event yield in a given $(p_{miss}, E_{miss})$ bin as:
\begin{equation}
  \frac{d^{6}\sigma(p_{miss}, E_{miss})}{dE_{e}dE_{p}d\Omega_{e}d\Omega_p} = \frac{Yield(p_{miss}, E_{miss})}{C \cdot t \cdot (\rho/A) \cdot b \cdot  V_B \cdot C_{Rad} \cdot C_{BM}},
\label{Eq:xsection}
  \end{equation}
where $C$ is the total accumulated beam charge, 
$t$ is the live time fraction in which the detectors are able to collect data, 
$A=3$ is the target atomic mass,  
$\rho$ is the nominal areal density of the gas in the target cell, 
and $b$ is a  correction factor to account for changes in the target 
density caused by local beam heating.  
$b$ was determined by measuring the beam current dependence
of the inclusive event yield~\cite{Santiesteban:2018qwi}. 
$V_B$ is a factor that accounts for the detection phase space and acceptance correction
for the given $(p_{miss}, E_{miss})$ bin 
and $C_{Rad}$ and $C_{BM}$ are the radiative and bin migration
corrections, respectively.  
The \trit{} event yield was also corrected for the radioactive decay of 
$2.78 \pm 0.18\%$ of the target \trits nuclei to \hets in the six months 
between when the target was filled and when the experiment was conducted.

We used the SIMC \cite{Simc} spectrometer simulation package to simulate our experiment to calculate the $V_B$, $C_{Rad}$ and $C_{BM}$ terms in Eq.~\ref{Eq:xsection},
and to compare the measured cross section with theoretical calculations.
SIMC generates \eep events with the addition of radiation effects over a wide 
phase-space, propagates the generated events through a spectrometer model to
account for acceptance and resolution effects, and then weights each accepted event
by a model cross section calculated for the original kinematics of that specific event.  
The weighted events are subsequently analyzed as the data and can be used to 
compare between the data and different model cross-section predictions.

We considered two PWIA cross-section models: (1) Faddeev-formulation-based
calculations by J. Golak et al.~\cite{CARASCO200341,BERMUTH2003199,Golak:2005iy} that either include or exclude the continuum interaction between the two
spectator nucleons (FSI$_{23}$), labeled Cracow  and Cracow-PW respectively, and (2) a factorized calculation using the \het{} spectral function of C. Ciofi degli Atti
and L. P. Kaptari including FSI$_{23}$ \cite{CiofidegliAtti:2004jg}
and the $\sigma_{cc1}$ electron off-shell nucleon
cross section~\cite{DeForest:1983ahx}, labeled CK+$CC1$.  
Due to the lack of \trit{} proton spectral functions, we assumed isospin 
symmetry and used the \het{} neutron spectral function for the \trit\eep simulation.
In addition, the Cracow calculation used the CD-Bonn nucleon-nucleon
potential \cite{Machleidt:2000ge} and CK used AV18
\cite{Wiringa:1994wb}.  To make consistent comparisons within this
work, we rescaled the CK calculation for each nucleus by the ratio of
the proton momentum distribution obtained with CD-Bonn relative to
that obtained with AV18 based on calculations in
Ref.~\cite{Marcucci:2018llz}.

\begin{figure}[t]
\includegraphics[width = 8.5cm, page=1]{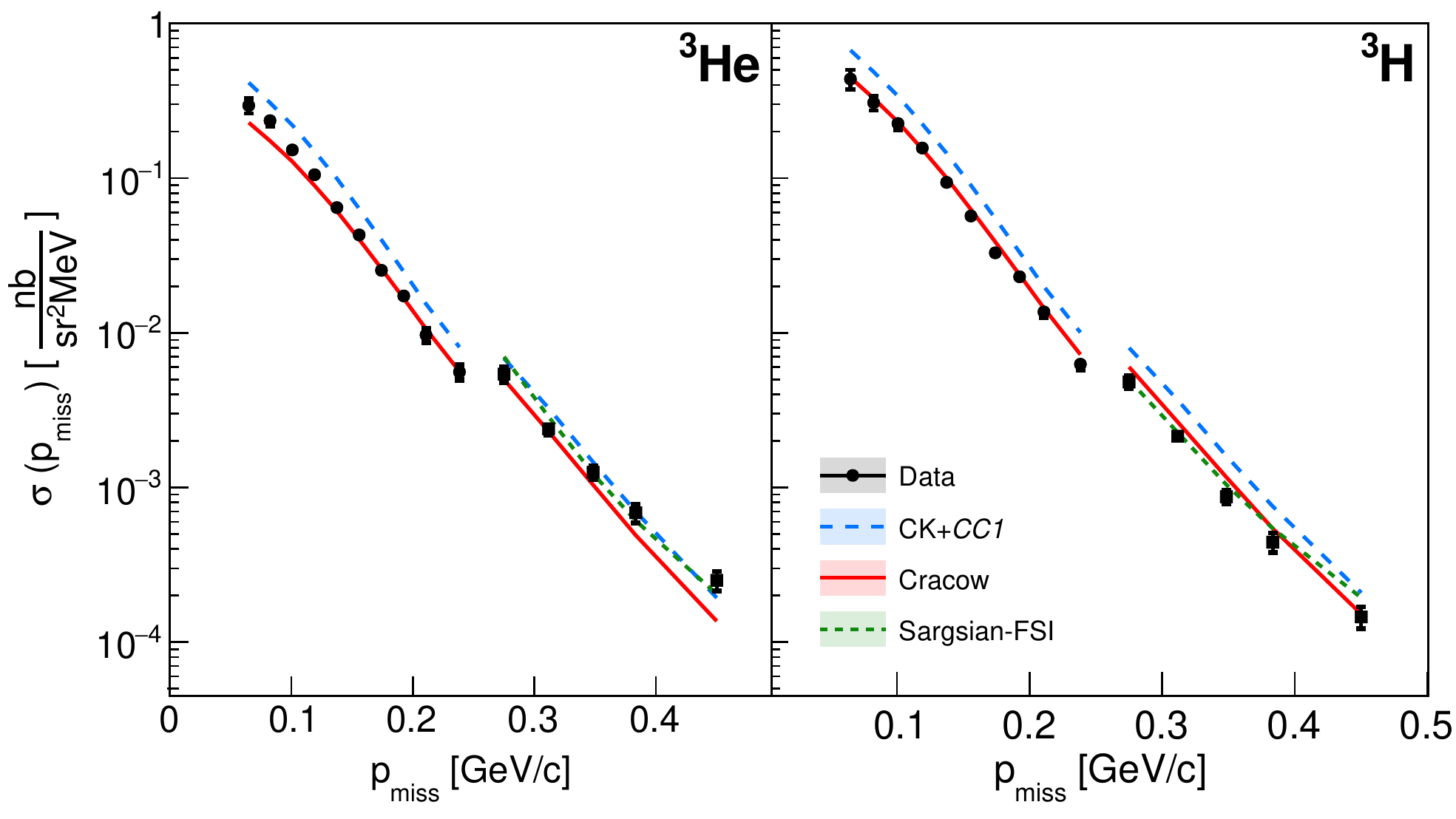}
\caption{ Absolute cross section as a function of \pmiss{} for $^3$He
  (left) and $^3$H (right).  The different sets of data points, depicted by black circles and squares, 
  correspond to the cross sections measured in the low-\pmiss{} and high-\pmiss{}
  kinematical settings respectively.  The lines correspond to cross
  sections calculated from different theoretical models, Cracow (solid
  red), CK+$CC1$ (dashed blue) and Sargsian-FSI (dotted
  green, $\pmiss>250$ MeV/c only).
  The different kinematical settings have different average elementary electron-nucleon
  cross-sections and therefore have a different overall scale for both data and calculations.}
\label{fig:AbsCross}
\end{figure}

\begin{figure*}[t]
\includegraphics[width = 6.5cm, page=2]{XSection.pdf}
\includegraphics[width = 6.5cm, page=1]{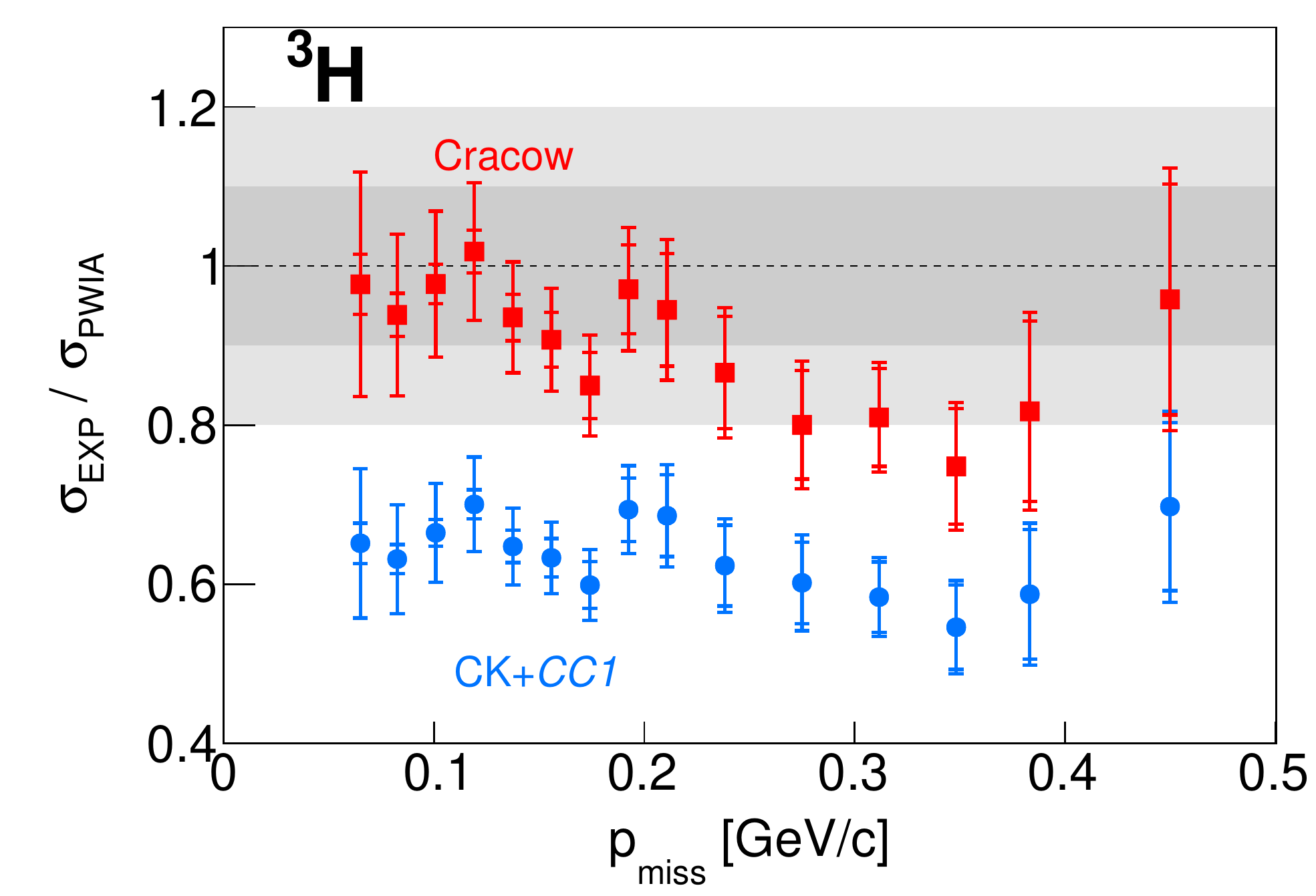}
\caption{The ratio of the experimental cross section to different PWIA
  calculations plotted versus \pmiss{} for \het\eep (top) 
  and \trit\eep (bottom).  Red squares show the ratio to the
  Cracow calculation while blue circles show the ratio to the 
  Ciofi-Kaptari spectral-function-based calculations (CK+$CC1$) (see text for details). Open symbols
  show the \het\eep\ data of Ref.~\cite{Benmokhtar:2004fs}, taken at lower $Q^2$ and
  $x\sim1$ kinematics, compared with the PWIA calculations of
  Ref.~\cite{CiofidegliAtti:2005qt,Laget:2004sm,Frankfurt:2008zv,Alvioli:2009zy}.
  The inner and outer bars show the statistical and statistical plus
  systematic uncertainties respectively.
  The shaded regions show $10\%$ and $20\%$ agreement intervals.}
\label{fig:resultsPWIA}
\end{figure*}

We corrected the \het{} and \trit{} cross sections for radiation and
bin migration effects using SIMC and the CK+$CC1$ cross-section model
that reproduces the \pmiss\ dependence of the measured cross section well.
Due to the excellent resolution of the HRS, bin migration effects were very small.  
Radiation effects were also small for \trit{} ($\lesssim 20\%$), but significant for
\het{} at low-\pmiss\ due to two-body breakup events that reconstructed to 
$\emiss > 8$ MeV due to radiation.  Since the \het{} cross section at high \emiss{} is dominated by radiative effects, we required $\emiss<50$ and 80 MeV for the 
low-  and high-\pmiss{} kinematics respectively.  

We then integrated the two dimensional experimental and theoretical cross
sections, $\sigma(\pmiss,\emiss)$, over \emiss{} to get the cross sections as 
a function of \pmiss. 

To facilitate comparison with future theoretical calculations, we
bin-centered the resulting cross sections, using the ratio of the
point theoretical cross section to the acceptance-averaged theoretical
cross section.  We calculated the point theoretical cross section by
summing the cross section evaluated at the central ($\langle Q^2 \rangle ,\langle x_B \rangle$) 
values over the seven \emiss-bins for that \pmiss{} as follows:
\begin{equation}
\begin{split}
&\sigma_{point}(\pmiss)= \\
&\sum_{j=1}^{7}\sigma(\langle Q^2\rangle ^j,
\langle x_B\rangle ^j,\pmiss,\emiss^j) \times \Delta\emiss^{j}
\label{eq:pointsigma}
\end{split}
\end{equation}
where $j$ labels the \emiss{} bin and $\Delta\emiss^{j}$ is the bin width.
We used both the Cracow and CK+$CC1$ cross-section models 
for this calculation, taking their average as the correction factor and
their difference divided by $\sqrt{12}$ as a measure of its $1\sigma$ uncertainty.  Future calculations can 
directly compare to our data by calculating the cross section at a small number 
of points and using Eq. \ref{eq:pointsigma}, rather than by computationally-intensive
integration over spectrometer acceptances. 

The point-to-point systematical uncertainties due to the event
selection criteria (momentum and angular acceptances, and
$\theta_{rq}$ and $x_B$ limits) were determined by repeating the
analysis 100 times, selecting each criterion randomly within
reasonable limits for each iteration. The systematic uncertainty was
taken to be the standard deviation of the resulting distribution cross sections.
They range from 1\% to 8\% and are typically much smaller than the statistical 
uncertainties.  Additional point-to-point systematics are due to bin-migration, 
bin-centering and radiative corrections and range between 0.5\% and 3.5\%.  

The overall normalization uncertainty of our measurement equals $2.7\%$,
and is due to uncertainty in the target density (1.5\%), beam-charge measurement 
run-by-run stability ($1\%$), tritium decay correction (0.15\%), and spectrometer detection and trigger efficiencies (2\%).

For completeness we also used SIMC to calculate the acceptance-averaged
cross sections using both Cracow and CK+$CC1$
cross-section models and compared them to our measured data before any
bin-centering corrections.  Both models well reproduce the shape of
the measured \emiss\ and \pmiss\ event distributions.
The ratio of the acceptance-averaged experimental to theoretical cross section is similar to the bin-centered ratios shown here.


Fig.~\ref{fig:AbsCross} shows the experimental, bin-centered, \het{}
and \trit\eep cross sections as a function of \pmiss{} and integrated over \emiss{}
from $8$ to $50$ or $80$ MeV for the low- and high-\pmiss{}
kinematics, respectively.  The cross section drops more than a factor
of $10^3$ from the lowest to highest \pmiss.  The Cracow calculation
appears to agree well with measured cross sections for \het{} for $\pmiss < 350$
MeV/c and for \trit{} at all \pmiss, while the CK+$CC1$ calculation
generally overestimates the measured cross sections.

For ease of comparison, Fig.~\ref{fig:resultsPWIA} shows the same measured cross sections
divided by the PWIA calculations. For \trit, the Cracow calculation
agrees with the data to about $20\%$.  For \het, the two agree for 
$100 \le \pmiss \le \SI{350}{\mega\eVperc}$ but disagree
by up to a factor of two for larger and lower \pmiss.  For both nuclei
the CK+$CC1$ calculation is higher than the data by about $60\%$.
These results are consistent with our \het/\trit{} cross-section ratio extracted
from the same data \cite{Cruz-Torres:2019bqw}, which agreed
with ratios of cross-section calculations and ratios of ground-state momentum distributions up to
$\pmiss \approx \SI{350}{\mega\eVperc}$.  The unexpected increase in the
\het/\trit{} cross section ratio at larger
\pmiss{} now appears to be due to both a decrease in the \trit\eep and an
increase in the \het\eep cross sections, relative to PWIA calculations.
As explained below, our data suggests that this effect is due to SCX effects.

\begin{figure*}
\includegraphics[width = 6.5cm, page=3]{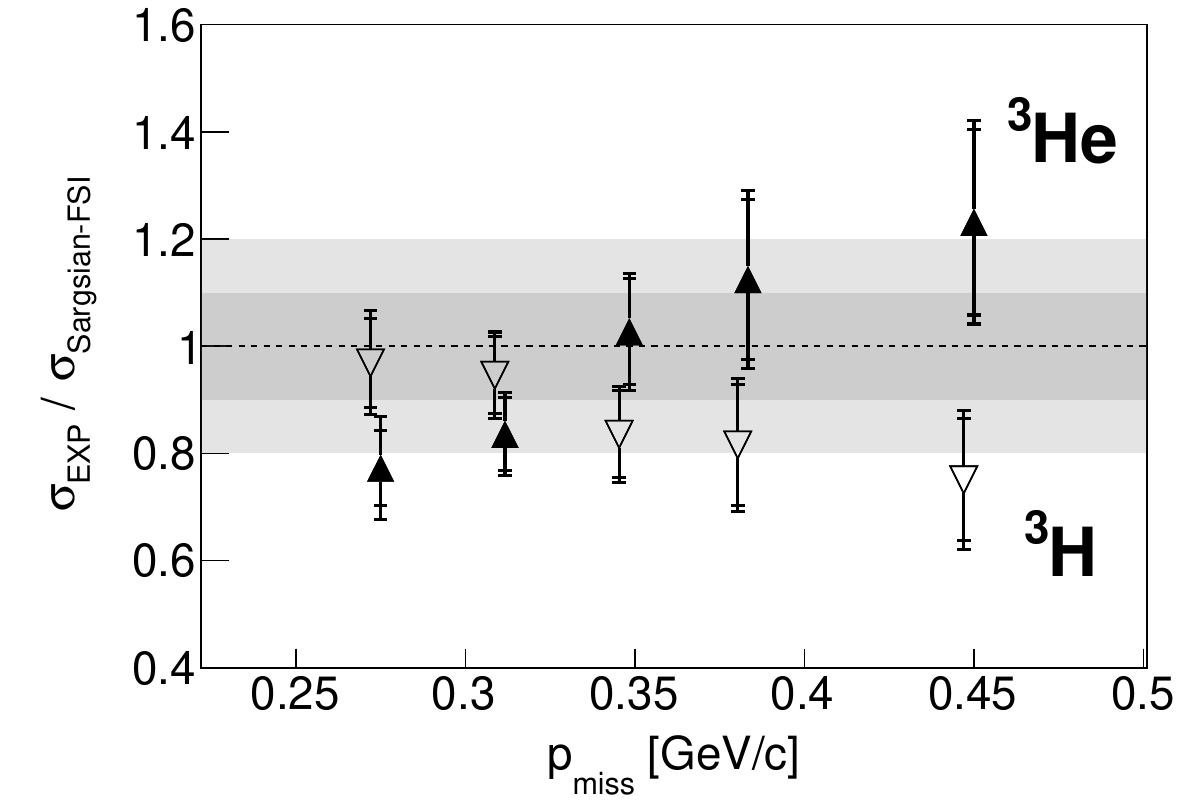}
\includegraphics[width = 6.5cm, page=2]{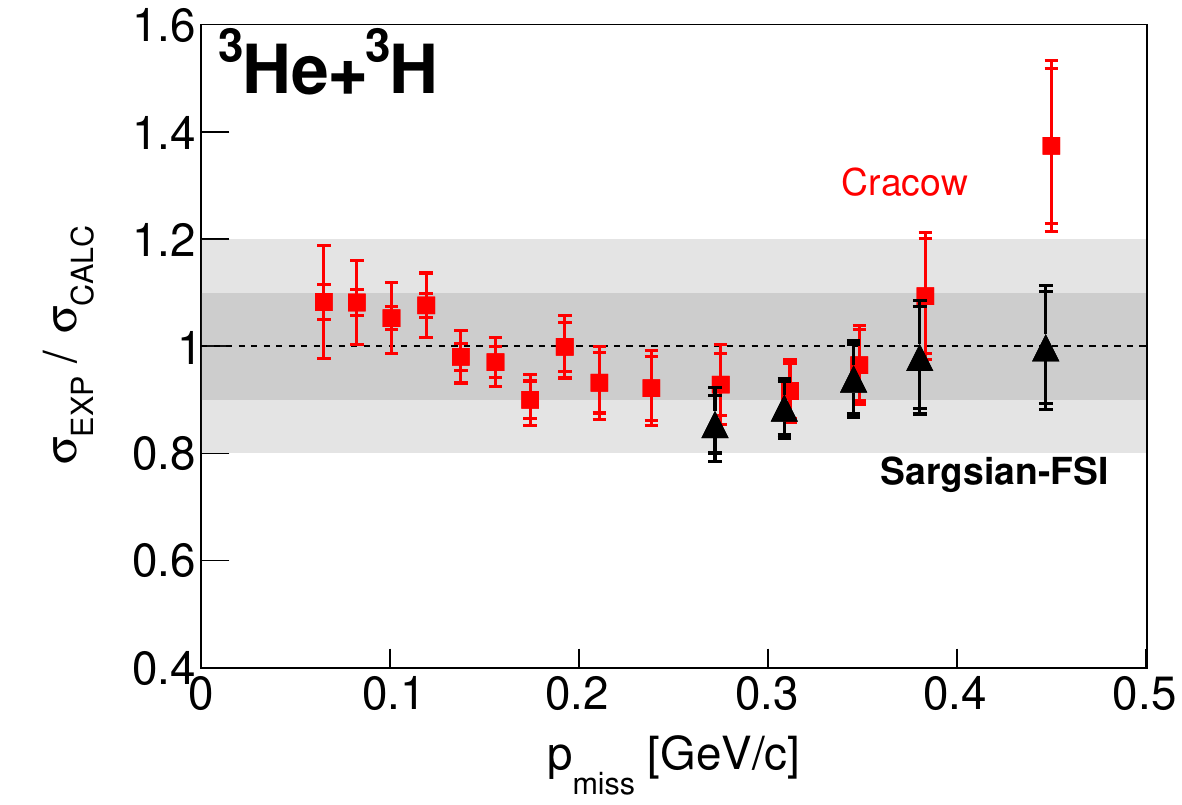}
\caption{Top: The ratio of the experimental cross sections to the
  calculation of Sargsian that includes FSI of the leading nucleon for
  \het{} (filled upright triangles) and \trit{} (open inverted triangles).
  Bottom: the ratio of the measured total \het{} + \trit{} cross section to the Cracow PWIA 
  calculation (red squares) and the Sargsian calculation that includes
  FSI (black triangles).  The inner and outer bars show the statistical and statistical plus
  systematic uncertainties respectively.
On both panels the shaded regions show $10\%$ and $20\%$ agreement intervals.}
\label{fig:resultsFSI}
\end{figure*}

The most recent \het\eep
three-body breakup cross-section measurements were done at $Q^2 = 1.5$ (GeV$/c$)$^2$
and $x_B = 1$~\cite{Benmokhtar:2004fs}, near the expected maximum of
struck-proton rescattering.  The measured cross-sections were lower
than PWIA calculations by a factor of $\sim 2$ for $\pmiss < 250$ MeV/c 
and higher by a factor of $\sim 3$ for $400 < \pmiss < 500$ MeV/c 
(see Fig.~\ref{fig:resultsPWIA}).  
These deviations were described by calculations which included 
the contribution of non-QE reaction mechanisms, primarily
FSI~\cite{CiofidegliAtti:2005qt,Laget:2004sm,Frankfurt:2008zv,Alvioli:2009zy}.
The large contribution of such non-QE reaction mechanisms to the
measured \eep cross-sections significantly limited their ability to constrain the
nucleon distributions at high momenta.  These non-QE effects are much
smaller in the current measurement due to our choice of kinematics.

In order to estimate the effects of struck-proton rescattering, we
also considered a calculation by M. Sargsian~\cite{SargsPRivate} which
accounts for the FSI of the struck-nucleon using the generalized
Eikonal approximation~\cite{misak05a,misak05b}, following the initial
PWIA proton knockout.  This calculation does not include the continuum
interaction between the two spectator nucleons, FSI$_{23}$, and is
therefore only applicable where those effects are small.  Comparing
the Cracow calculations with and without FSI$_{23}$ showed that its
effects decrease rapidly with
\pmiss. 
We therefore used the Sargsian FSI calculations only at $\pmiss\ge
250$ MeV/c.  We further verified that using this model for bin
centering did not significantly change the correction factors.

See the online supplemental materials for more details on the
kinematics, analysis procedures, and theoretical corrections, and for
tables of all measured and calculated cross sections.

Fig.~\ref{fig:resultsFSI} (top) shows the ratio of the experimental, bin-centered
cross-section to the Sargsian FSI calculation for $\pmiss > 250$
MeV/c.  The FSI calculation generally agrees with the data.  The 
trend of the ratio seems to be opposite for \het{} and \trit{} with
the former rising above unity while the latter decreasing below it.

If the high-momentum proton and neutron densities are equal in both
\het{} and \trit{}, 
this trend could result from SCX processes which would increase 
the \het\eep cross section, but decrease the \trit\eep cross section.
While further calculations are needed to fully quantify this effect,
this equal-density assumption is supported 
both by ab-initio calculations~\cite{Wiringa:2014} and by previous measurements
that showed that, at high \pmiss, electrons scatter primarily off
nucleons in $np$-short-range correlated pairs~\cite{Hen:2016kwk,Atti:2015eda,piasetzky06,subedi08,korover14,Hen:2014nza,Cohen:2018gzh,Duer:2018sby,Duer:2018sxh,Pybus:2020itv,schmidt20}.
We can test this, since SCX effects should be suppressed in isoscalar systems
due to large cancellations between $(n,p)$ and $(p,n)$ processes. 
This implies that the isoscalar $A=3$ cross-section (i.e.,  \het{} + \trit{})
is insensitive to these effects.

The ratio of the measured total isoscalar $A=3$ cross-section of
\het{} + \trit{} to the Cracow and Sargsian calculations is shown in
Fig.~\ref{fig:resultsFSI} (bottom). As expected, both calculations
agree with the data to within about $\pm10\%$, comparable to the accuracy
of the data.  Due to the QE nature of our measurement, this excellent agreement 
between our isoscalar data and ab-initio nuclear theory validates calculations of the $A=3$ ground
state momentum distribution up to extremely high nucleon momenta of about 500~MeV$/c$.


To conclude, we present new \het{} and \trit\eep cross-section
measurements, which represent the first new high-energy electron
scattering data on tritium in over 30 years. By choosing kinematics
specifically to minimize non-quasielastic contributions (high-$Q^2$,
$x_B>1$, $\theta_{rq}<37.5^\circ$), the
data are much more directly sensitive to the properties of the $A=3$
nuclear ground state.  PWIA calculations can reproduce the \het{} data
to within $\pm20\%$ for $100 \le \pmiss \le 350$ MeV/c, a significant improvement over previous
measurements at $x_B=1$, and do even better for \trit, where they
can reproduce the data to  $\pm20\%$ over the entire measured \pmiss
range. A calculation that includes leading nucleon rescattering
improves agreement at high \pmiss, and the residual
disagreement has the same sign as would be expected from additional
SCX contributions. The isoscalar (\het{} + \trit) cross-section agrees
remarkably well with QE cross-section calculations, validating both the choice of 
kinematics and calculations of the $A=3$ ground state up to extremely
high nucleon momenta of 500 MeV/c.



These data are a crucial benchmark for few-body nuclear theory
and are a necessary, but not sufficient, test of theoretical calculations that are also used in the study
of heavier nuclear systems.

\begin{acknowledgements}
We acknowledge the contribution of the Jefferson-Lab target group and
technical staff for design and construction of the tritium target and
their support running this experiment. We thank C. Ciofi degli Atti
and L. Kaptari for the $^3$He spectral function calculations and  M. Sargsian for the FSI calculations.
We also thank M. Strikman for many valuable discussions. 
This work was supported by the U.S. Department of Energy (DOE) grant DE-AC05-06OR23177 under which Jefferson Science Associates, LLC, operates the Thomas Jefferson National Accelerator Facility, 
the U.S. National Science Foundation,
the Pazi foundation, and the Israel Science Foundation.  
The Kent State University contribution is supported under the PHY-1714809 grant from the U.S. National Science Foundation.
The University of Tennessee contribution is supported by the DE-SC0013615 grant.
The work of ANL group members is supported by DOE grant DE-AC02-06CH11357.
The contribution of the Cracow group was supported by the Polish National
Science Centre
under Grants No. 2016/22/M/ST2/00173 and No.2016/21/D/ST2/01120.
The numerical calculations were partially performed on the supercomputer
cluster of the JSC, J\"ulich, Germany.
The Temple University group is supported by the DOE award DE-SC0016577.

\end{acknowledgements}

\bibliography{../../TritiumBib2}

\begin{thebibliography}{63}%
\makeatletter
\providecommand \@ifxundefined [1]{%
 \@ifx{#1\undefined}
}%
\providecommand \@ifnum [1]{%
 \ifnum #1\expandafter \@firstoftwo
 \else \expandafter \@secondoftwo
 \fi
}%
\providecommand \@ifx [1]{%
 \ifx #1\expandafter \@firstoftwo
 \else \expandafter \@secondoftwo
 \fi
}%
\providecommand \natexlab [1]{#1}%
\providecommand \enquote  [1]{``#1''}%
\providecommand \bibnamefont  [1]{#1}%
\providecommand \bibfnamefont [1]{#1}%
\providecommand \citenamefont [1]{#1}%
\providecommand \href@noop [0]{\@secondoftwo}%
\providecommand \href [0]{\begingroup \@sanitize@url \@href}%
\providecommand \@href[1]{\@@startlink{#1}\@@href}%
\providecommand \@@href[1]{\endgroup#1\@@endlink}%
\providecommand \@sanitize@url [0]{\catcode `\\12\catcode `\$12\catcode
  `\&12\catcode `\#12\catcode `\^12\catcode `\_12\catcode `\%12\relax}%
\providecommand \@@startlink[1]{}%
\providecommand \@@endlink[0]{}%
\providecommand \url  [0]{\begingroup\@sanitize@url \@url }%
\providecommand \@url [1]{\endgroup\@href {#1}{\urlprefix }}%
\providecommand \urlprefix  [0]{URL }%
\providecommand \Eprint [0]{\href }%
\providecommand \doibase [0]{http://dx.doi.org/}%
\providecommand \selectlanguage [0]{\@gobble}%
\providecommand \bibinfo  [0]{\@secondoftwo}%
\providecommand \bibfield  [0]{\@secondoftwo}%
\providecommand \translation [1]{[#1]}%
\providecommand \BibitemOpen [0]{}%
\providecommand \bibitemStop [0]{}%
\providecommand \bibitemNoStop [0]{.\EOS\space}%
\providecommand \EOS [0]{\spacefactor3000\relax}%
\providecommand \BibitemShut  [1]{\csname bibitem#1\endcsname}%
\let\auto@bib@innerbib\@empty
\bibitem [{\citenamefont {Kelly}(1996)}]{Kelly:1996hd}%
  \BibitemOpen
  \bibfield  {author} {\bibinfo {author} {\bibfnamefont {J.~J.}\ \bibnamefont
  {Kelly}},\ }\href {\doibase 10.1007/0-306-47067-5_2} {\bibfield  {journal}
  {\bibinfo  {journal} {Adv. Nucl. Phys.}\ }\textbf {\bibinfo {volume} {23}},\
  \bibinfo {pages} {75} (\bibinfo {year} {1996})}\BibitemShut {NoStop}%
\bibitem [{\citenamefont {Ford}\ \emph {et~al.}(2014)\citenamefont {Ford},
  \citenamefont {Jeschonnek},\ and\ \citenamefont {Van~Orden}}]{Ford:2014yua}%
  \BibitemOpen
  \bibfield  {author} {\bibinfo {author} {\bibfnamefont {W.~P.}\ \bibnamefont
  {Ford}}, \bibinfo {author} {\bibfnamefont {S.}~\bibnamefont {Jeschonnek}}, \
  and\ \bibinfo {author} {\bibfnamefont {J.~W.}\ \bibnamefont {Van~Orden}},\
  }\href {\doibase 10.1103/PhysRevC.90.064006} {\bibfield  {journal} {\bibinfo
  {journal} {Phys. Rev.}\ }\textbf {\bibinfo {volume} {C90}},\ \bibinfo {pages}
  {064006} (\bibinfo {year} {2014})},\ \Eprint {http://arxiv.org/abs/1411.3306}
  {arXiv:1411.3306 [nucl-th]} \BibitemShut {NoStop}%
\bibitem [{\citenamefont {Golak}\ \emph {et~al.}(2005)\citenamefont {Golak},
  \citenamefont {Skibi\'nski}, \citenamefont {Wita{\l}a}, \citenamefont
  {Gl\"ockle}, \citenamefont {Nogga},\ and\ \citenamefont
  {Kamada}}]{Golak:2005iy}%
  \BibitemOpen
  \bibfield  {author} {\bibinfo {author} {\bibfnamefont {J.}~\bibnamefont
  {Golak}}, \bibinfo {author} {\bibfnamefont {R.}~\bibnamefont {Skibi\'nski}},
  \bibinfo {author} {\bibfnamefont {H.}~\bibnamefont {Wita{\l}a}}, \bibinfo
  {author} {\bibfnamefont {W.}~\bibnamefont {Gl\"ockle}}, \bibinfo {author}
  {\bibfnamefont {A.}~\bibnamefont {Nogga}}, \ and\ \bibinfo {author}
  {\bibfnamefont {H.}~\bibnamefont {Kamada}},\ }\href {\doibase
  10.1016/j.physrep.2005.04.005} {\bibfield  {journal} {\bibinfo  {journal}
  {Phys. Rept.}\ }\textbf {\bibinfo {volume} {415}},\ \bibinfo {pages} {89}
  (\bibinfo {year} {2005})},\ \Eprint {http://arxiv.org/abs/nucl-th/0505072}
  {arXiv:nucl-th/0505072 [nucl-th]} \BibitemShut {NoStop}%
\bibitem [{\citenamefont {Sick}(2001)}]{Sick:2001rh}%
  \BibitemOpen
  \bibfield  {author} {\bibinfo {author} {\bibfnamefont {I.}~\bibnamefont
  {Sick}},\ }\href {\doibase 10.1016/S0146-6410(01)00156-9} {\bibfield
  {journal} {\bibinfo  {journal} {Prog. Part. Nucl. Phys.}\ }\textbf {\bibinfo
  {volume} {47}},\ \bibinfo {pages} {245} (\bibinfo {year} {2001})},\ \Eprint
  {http://arxiv.org/abs/nucl-ex/0208009} {arXiv:nucl-ex/0208009 [nucl-ex]}
  \BibitemShut {NoStop}%
\bibitem [{\citenamefont {Benmokhtar}\ \emph {et~al.}(2005)\citenamefont
  {Benmokhtar} \emph {et~al.}}]{Benmokhtar:2004fs}%
  \BibitemOpen
  \bibfield  {author} {\bibinfo {author} {\bibfnamefont {F.}~\bibnamefont
  {Benmokhtar}} \emph {et~al.} (\bibinfo {collaboration} {Jefferson Lab Hall
  A}),\ }\href {\doibase 10.1103/PhysRevLett.94.082305} {\bibfield  {journal}
  {\bibinfo  {journal} {Phys. Rev. Lett.}\ }\textbf {\bibinfo {volume} {94}},\
  \bibinfo {pages} {082305} (\bibinfo {year} {2005})},\ \Eprint
  {http://arxiv.org/abs/nucl-ex/0408015} {arXiv:nucl-ex/0408015 [nucl-ex]}
  \BibitemShut {NoStop}%
\bibitem [{\citenamefont {Rvachev}\ \emph {et~al.}(2005)\citenamefont {Rvachev}
  \emph {et~al.}}]{Rvachev:2004yr}%
  \BibitemOpen
  \bibfield  {author} {\bibinfo {author} {\bibfnamefont {M.~M.}\ \bibnamefont
  {Rvachev}} \emph {et~al.} (\bibinfo {collaboration} {Jefferson Lab Hall A}),\
  }\href {\doibase 10.1103/PhysRevLett.94.192302} {\bibfield  {journal}
  {\bibinfo  {journal} {Phys. Rev. Lett.}\ }\textbf {\bibinfo {volume} {94}},\
  \bibinfo {pages} {192302} (\bibinfo {year} {2005})},\ \Eprint
  {http://arxiv.org/abs/nucl-ex/0409005} {arXiv:nucl-ex/0409005 [nucl-ex]}
  \BibitemShut {NoStop}%
\bibitem [{\citenamefont {Long}\ \emph {et~al.}(2019)\citenamefont {Long} \emph
  {et~al.}}]{Long:2019iig}%
  \BibitemOpen
  \bibfield  {author} {\bibinfo {author} {\bibfnamefont {E.}~\bibnamefont
  {Long}} \emph {et~al.},\ }\href {\doibase 10.1016/j.physletb.2019.134875}
  {\bibfield  {journal} {\bibinfo  {journal} {Phys. Lett.}\ }\textbf {\bibinfo
  {volume} {B797}},\ \bibinfo {pages} {134875} (\bibinfo {year} {2019})},\
  \Eprint {http://arxiv.org/abs/1906.04075} {arXiv:1906.04075 [nucl-ex]}
  \BibitemShut {NoStop}%
\bibitem [{\citenamefont {Mihovilovic}\ \emph {et~al.}(2014)\citenamefont
  {Mihovilovic} \emph {et~al.}}]{Mihovilovic:2014gdi}%
  \BibitemOpen
  \bibfield  {author} {\bibinfo {author} {\bibfnamefont {M.}~\bibnamefont
  {Mihovilovic}} \emph {et~al.} (\bibinfo {collaboration} {Jefferson Lab Hall
  A}),\ }\href {\doibase 10.1103/PhysRevLett.113.232505} {\bibfield  {journal}
  {\bibinfo  {journal} {Phys. Rev. Lett.}\ }\textbf {\bibinfo {volume} {113}},\
  \bibinfo {pages} {232505} (\bibinfo {year} {2014})},\ \Eprint
  {http://arxiv.org/abs/1409.2253} {arXiv:1409.2253 [nucl-ex]} \BibitemShut
  {NoStop}%
\bibitem [{\citenamefont {Mihovilovič}\ \emph {et~al.}(2019)\citenamefont
  {Mihovilovič} \emph {et~al.}}]{Mihovilovic:2018fux}%
  \BibitemOpen
  \bibfield  {author} {\bibinfo {author} {\bibfnamefont {M.}~\bibnamefont
  {Mihovilovič}} \emph {et~al.} (\bibinfo {collaboration} {Jefferson Lab Hall
  A}),\ }\href {\doibase 10.1016/j.physletb.2018.10.063} {\bibfield  {journal}
  {\bibinfo  {journal} {Phys. Lett.}\ }\textbf {\bibinfo {volume} {B788}},\
  \bibinfo {pages} {117} (\bibinfo {year} {2019})},\ \Eprint
  {http://arxiv.org/abs/1804.06043} {arXiv:1804.06043 [nucl-ex]} \BibitemShut
  {NoStop}%
\bibitem [{\citenamefont {Zhang}\ \emph {et~al.}(2015)\citenamefont {Zhang}
  \emph {et~al.}}]{Zhang:2015kna}%
  \BibitemOpen
  \bibfield  {author} {\bibinfo {author} {\bibfnamefont {Y.~W.}\ \bibnamefont
  {Zhang}} \emph {et~al.},\ }\href {\doibase 10.1103/PhysRevLett.115.172502}
  {\bibfield  {journal} {\bibinfo  {journal} {Phys. Rev. Lett.}\ }\textbf
  {\bibinfo {volume} {115}},\ \bibinfo {pages} {172502} (\bibinfo {year}
  {2015})},\ \Eprint {http://arxiv.org/abs/1502.02636} {arXiv:1502.02636
  [nucl-ex]} \BibitemShut {NoStop}%
\bibitem [{\citenamefont {Camsonne}\ \emph {et~al.}(2017)\citenamefont
  {Camsonne} \emph {et~al.}}]{Camsonne:2016ged}%
  \BibitemOpen
  \bibfield  {author} {\bibinfo {author} {\bibfnamefont {A.}~\bibnamefont
  {Camsonne}} \emph {et~al.},\ }\href {\doibase 10.1103/PhysRevLett.119.209901,
  10.1103/PhysRevLett.119.162501} {\bibfield  {journal} {\bibinfo  {journal}
  {Phys. Rev. Lett.}\ }\textbf {\bibinfo {volume} {119}},\ \bibinfo {pages}
  {162501} (\bibinfo {year} {2017})},\ \bibinfo {note} {[Addendum: Phys. Rev.
  Lett.119,no.20,209901(2017)]},\ \Eprint {http://arxiv.org/abs/1610.07456}
  {arXiv:1610.07456 [nucl-ex]} \BibitemShut {NoStop}%
\bibitem [{\citenamefont {Riordan}\ \emph {et~al.}(2010)\citenamefont {Riordan}
  \emph {et~al.}}]{Riordan:2010id}%
  \BibitemOpen
  \bibfield  {author} {\bibinfo {author} {\bibfnamefont {S.}~\bibnamefont
  {Riordan}} \emph {et~al.},\ }\href {\doibase 10.1103/PhysRevLett.105.262302}
  {\bibfield  {journal} {\bibinfo  {journal} {Phys. Rev. Lett.}\ }\textbf
  {\bibinfo {volume} {105}},\ \bibinfo {pages} {262302} (\bibinfo {year}
  {2010})},\ \Eprint {http://arxiv.org/abs/1008.1738} {arXiv:1008.1738
  [nucl-ex]} \BibitemShut {NoStop}%
\bibitem [{\citenamefont {Collard}\ \emph {et~al.}(1963)\citenamefont
  {Collard}, \citenamefont {Hofstadter}, \citenamefont {Johansson},
  \citenamefont {Parks}, \citenamefont {Ryneveld}, \citenamefont {Walker},
  \citenamefont {Yearian}, \citenamefont {Day},\ and\ \citenamefont
  {Wagner}}]{Collard:1963zza}%
  \BibitemOpen
  \bibfield  {author} {\bibinfo {author} {\bibfnamefont {H.}~\bibnamefont
  {Collard}}, \bibinfo {author} {\bibfnamefont {R.}~\bibnamefont {Hofstadter}},
  \bibinfo {author} {\bibfnamefont {A.}~\bibnamefont {Johansson}}, \bibinfo
  {author} {\bibfnamefont {R.}~\bibnamefont {Parks}}, \bibinfo {author}
  {\bibfnamefont {M.}~\bibnamefont {Ryneveld}}, \bibinfo {author}
  {\bibfnamefont {A.}~\bibnamefont {Walker}}, \bibinfo {author} {\bibfnamefont
  {M.~R.}\ \bibnamefont {Yearian}}, \bibinfo {author} {\bibfnamefont {R.~B.}\
  \bibnamefont {Day}}, \ and\ \bibinfo {author} {\bibfnamefont {R.~T.}\
  \bibnamefont {Wagner}},\ }\href {\doibase 10.1103/PhysRevLett.11.132}
  {\bibfield  {journal} {\bibinfo  {journal} {Phys. Rev. Lett.}\ }\textbf
  {\bibinfo {volume} {11}},\ \bibinfo {pages} {132} (\bibinfo {year}
  {1963})}\BibitemShut {NoStop}%
\bibitem [{\citenamefont {Schiff}\ \emph {et~al.}(1963)\citenamefont {Schiff},
  \citenamefont {Collard}, \citenamefont {Hofstadter}, \citenamefont
  {Johansson},\ and\ \citenamefont {Yearian}}]{Schiff:1963zzb}%
  \BibitemOpen
  \bibfield  {author} {\bibinfo {author} {\bibfnamefont {L.~I.}\ \bibnamefont
  {Schiff}}, \bibinfo {author} {\bibfnamefont {H.}~\bibnamefont {Collard}},
  \bibinfo {author} {\bibfnamefont {R.}~\bibnamefont {Hofstadter}}, \bibinfo
  {author} {\bibfnamefont {A.}~\bibnamefont {Johansson}}, \ and\ \bibinfo
  {author} {\bibfnamefont {M.~R.}\ \bibnamefont {Yearian}},\ }\href {\doibase
  10.1103/PhysRevLett.11.387} {\bibfield  {journal} {\bibinfo  {journal} {Phys.
  Rev. Lett.}\ }\textbf {\bibinfo {volume} {11}},\ \bibinfo {pages} {387}
  (\bibinfo {year} {1963})}\BibitemShut {NoStop}%
\bibitem [{\citenamefont {Johansson}(1964)}]{PhysRev.136.B1030}%
  \BibitemOpen
  \bibfield  {author} {\bibinfo {author} {\bibfnamefont {A.}~\bibnamefont
  {Johansson}},\ }\href {\doibase 10.1103/PhysRev.136.B1030} {\bibfield
  {journal} {\bibinfo  {journal} {Phys. Rev.}\ }\textbf {\bibinfo {volume}
  {136}},\ \bibinfo {pages} {B1030} (\bibinfo {year} {1964})}\BibitemShut
  {NoStop}%
\bibitem [{\citenamefont {GRIFFY}\ and\ \citenamefont
  {OAKES}(1965)}]{RevModPhys.37.402}%
  \BibitemOpen
  \bibfield  {author} {\bibinfo {author} {\bibfnamefont {T.~A.}\ \bibnamefont
  {GRIFFY}}\ and\ \bibinfo {author} {\bibfnamefont {R.~J.}\ \bibnamefont
  {OAKES}},\ }\href {\doibase 10.1103/RevModPhys.37.402} {\bibfield  {journal}
  {\bibinfo  {journal} {Rev. Mod. Phys.}\ }\textbf {\bibinfo {volume} {37}},\
  \bibinfo {pages} {402} (\bibinfo {year} {1965})}\BibitemShut {NoStop}%
\bibitem [{\citenamefont {Dow}(1986)}]{Dow:1986auc}%
  \BibitemOpen
  \bibfield  {author} {\bibinfo {author} {\bibfnamefont {K.}~\bibnamefont
  {Dow}},\ }\bibfield  {booktitle} {\emph {\bibinfo {booktitle} {{Proceedings,
  Three Body Forces in the Three Nucleon System: Washington, DC, April 24-26,
  1986}}},\ }\href {\doibase 10.1007/3-540-16805-2_51} {\bibfield  {journal}
  {\bibinfo  {journal} {Lect. Notes Phys.}\ }\textbf {\bibinfo {volume}
  {260}},\ \bibinfo {pages} {346} (\bibinfo {year} {1986})}\BibitemShut
  {NoStop}%
\bibitem [{\citenamefont {Beck}(1986)}]{Beck:1986rhj}%
  \BibitemOpen
  \bibfield  {author} {\bibinfo {author} {\bibfnamefont {D.~H.}\ \bibnamefont
  {Beck}},\ }\bibfield  {booktitle} {\emph {\bibinfo {booktitle} {{Proceedings,
  Three Body Forces in the Three Nucleon System: Washington, DC, April 24-26,
  1986}}},\ }\href {\doibase 10.1007/3-540-16805-2_14} {\bibfield  {journal}
  {\bibinfo  {journal} {Lect. Notes Phys.}\ }\textbf {\bibinfo {volume}
  {260}},\ \bibinfo {pages} {138} (\bibinfo {year} {1986})}\BibitemShut
  {NoStop}%
\bibitem [{\citenamefont {Beck}\ \emph {et~al.}(1987)\citenamefont {Beck} \emph
  {et~al.}}]{Beck:1987zz}%
  \BibitemOpen
  \bibfield  {author} {\bibinfo {author} {\bibfnamefont {D.}~\bibnamefont
  {Beck}} \emph {et~al.},\ }\href {\doibase 10.1103/PhysRevLett.59.1537}
  {\bibfield  {journal} {\bibinfo  {journal} {Phys. Rev. Lett.}\ }\textbf
  {\bibinfo {volume} {59}},\ \bibinfo {pages} {1537} (\bibinfo {year}
  {1987})}\BibitemShut {NoStop}%
\bibitem [{\citenamefont {Dow}\ \emph {et~al.}(1988)\citenamefont {Dow} \emph
  {et~al.}}]{Dow:1988rk}%
  \BibitemOpen
  \bibfield  {author} {\bibinfo {author} {\bibfnamefont {K.}~\bibnamefont
  {Dow}} \emph {et~al.},\ }\href {\doibase 10.1103/PhysRevLett.61.1706}
  {\bibfield  {journal} {\bibinfo  {journal} {Phys. Rev. Lett.}\ }\textbf
  {\bibinfo {volume} {61}},\ \bibinfo {pages} {1706} (\bibinfo {year}
  {1988})}\BibitemShut {NoStop}%
\bibitem [{\citenamefont {Juster}\ \emph {et~al.}(1985)\citenamefont {Juster}
  \emph {et~al.}}]{Juster:1985sd}%
  \BibitemOpen
  \bibfield  {author} {\bibinfo {author} {\bibfnamefont {F.~P.}\ \bibnamefont
  {Juster}} \emph {et~al.},\ }\href {\doibase 10.1103/PhysRevLett.55.2261}
  {\bibfield  {journal} {\bibinfo  {journal} {Phys. Rev. Lett.}\ }\textbf
  {\bibinfo {volume} {55}},\ \bibinfo {pages} {2261} (\bibinfo {year}
  {1985})}\BibitemShut {NoStop}%
\bibitem [{\citenamefont {Amroun}\ \emph {et~al.}(1994)\citenamefont {Amroun}
  \emph {et~al.}}]{Amroun:1994qj}%
  \BibitemOpen
  \bibfield  {author} {\bibinfo {author} {\bibfnamefont {A.}~\bibnamefont
  {Amroun}} \emph {et~al.},\ }\href {\doibase 10.1016/0375-9474(94)90925-3}
  {\bibfield  {journal} {\bibinfo  {journal} {Nucl. Phys.}\ }\textbf {\bibinfo
  {volume} {A579}},\ \bibinfo {pages} {596} (\bibinfo {year}
  {1994})}\BibitemShut {NoStop}%
\bibitem [{\citenamefont {Alcorn}\ \emph {et~al.}(2004)\citenamefont {Alcorn}
  \emph {et~al.}}]{Alcorn:2004sb}%
  \BibitemOpen
  \bibfield  {author} {\bibinfo {author} {\bibfnamefont {J.}~\bibnamefont
  {Alcorn}} \emph {et~al.},\ }\href {\doibase 10.1016/j.nima.2003.11.415}
  {\bibfield  {journal} {\bibinfo  {journal} {Nucl. Instrum. Meth.}\ }\textbf
  {\bibinfo {volume} {A522}},\ \bibinfo {pages} {294} (\bibinfo {year}
  {2004})}\BibitemShut {NoStop}%
\bibitem [{\citenamefont {Santiesteban}\ \emph {et~al.}(2019)\citenamefont
  {Santiesteban} \emph {et~al.}}]{Santiesteban:2018qwi}%
  \BibitemOpen
  \bibfield  {author} {\bibinfo {author} {\bibfnamefont {S.~N.}\ \bibnamefont
  {Santiesteban}} \emph {et~al.},\ }\href {\doibase 10.1016/j.nima.2019.06.025}
  {\bibfield  {journal} {\bibinfo  {journal} {Nucl. Instrum. Meth.}\ }\textbf
  {\bibinfo {volume} {A940}},\ \bibinfo {pages} {351} (\bibinfo {year}
  {2019})},\ \Eprint {http://arxiv.org/abs/1811.12167} {arXiv:1811.12167
  [physics.ins-det]} \BibitemShut {NoStop}%
\bibitem [{\citenamefont {Gao}\ \emph {et~al.}(2000)\citenamefont {Gao} \emph
  {et~al.}}]{gao00}%
  \BibitemOpen
  \bibfield  {author} {\bibinfo {author} {\bibfnamefont {J.}~\bibnamefont
  {Gao}} \emph {et~al.} (\bibinfo {collaboration} {The Jefferson Lab Hall A
  Collaboration}),\ }\href {\doibase 10.1103/PhysRevLett.84.3265} {\bibfield
  {journal} {\bibinfo  {journal} {Phys. Rev. Lett.}\ }\textbf {\bibinfo
  {volume} {84}},\ \bibinfo {pages} {3265} (\bibinfo {year}
  {2000})}\BibitemShut {NoStop}%
\bibitem [{\citenamefont {Udias}\ \emph {et~al.}(1999)\citenamefont {Udias},
  \citenamefont {Caballero}, \citenamefont {Moya~de Guerra}, \citenamefont
  {Amaro},\ and\ \citenamefont {Donnelly}}]{udias99}%
  \BibitemOpen
  \bibfield  {author} {\bibinfo {author} {\bibfnamefont {J.~M.}\ \bibnamefont
  {Udias}}, \bibinfo {author} {\bibfnamefont {J.~A.}\ \bibnamefont
  {Caballero}}, \bibinfo {author} {\bibfnamefont {E.}~\bibnamefont {Moya~de
  Guerra}}, \bibinfo {author} {\bibfnamefont {J.~E.}\ \bibnamefont {Amaro}}, \
  and\ \bibinfo {author} {\bibfnamefont {T.~W.}\ \bibnamefont {Donnelly}},\
  }\href {\doibase 10.1103/PhysRevLett.83.5451} {\bibfield  {journal} {\bibinfo
   {journal} {Phys. Rev. Lett.}\ }\textbf {\bibinfo {volume} {83}},\ \bibinfo
  {pages} {5451} (\bibinfo {year} {1999})}\BibitemShut {NoStop}%
\bibitem [{\citenamefont {Alvarez-Rodriguez}\ \emph {et~al.}(2011)\citenamefont
  {Alvarez-Rodriguez}, \citenamefont {Udias}, \citenamefont {Vignote},
  \citenamefont {Garrido}, \citenamefont {Sarriguren}, \citenamefont {Moya~de
  Guerra}, \citenamefont {Pace}, \citenamefont {Kievsky},\ and\ \citenamefont
  {Salme}}]{AlvarezRodriguez:2010nb}%
  \BibitemOpen
  \bibfield  {author} {\bibinfo {author} {\bibfnamefont {R.}~\bibnamefont
  {Alvarez-Rodriguez}}, \bibinfo {author} {\bibfnamefont {J.~M.}\ \bibnamefont
  {Udias}}, \bibinfo {author} {\bibfnamefont {J.~R.}\ \bibnamefont {Vignote}},
  \bibinfo {author} {\bibfnamefont {E.}~\bibnamefont {Garrido}}, \bibinfo
  {author} {\bibfnamefont {P.}~\bibnamefont {Sarriguren}}, \bibinfo {author}
  {\bibfnamefont {E.}~\bibnamefont {Moya~de Guerra}}, \bibinfo {author}
  {\bibfnamefont {E.}~\bibnamefont {Pace}}, \bibinfo {author} {\bibfnamefont
  {A.}~\bibnamefont {Kievsky}}, \ and\ \bibinfo {author} {\bibfnamefont
  {G.}~\bibnamefont {Salme}},\ }\bibfield  {booktitle} {\emph {\bibinfo
  {booktitle} {{Proceedings, 21st European Conference on Few-Body Problems in
  Physics (EFB21): Salamanca, Castilla y Leon, Spain, August 29-September 3,
  2010}}},\ }\href {\doibase 10.1007/s00601-010-0187-4} {\bibfield  {journal}
  {\bibinfo  {journal} {Few Body Syst.}\ }\textbf {\bibinfo {volume} {50}},\
  \bibinfo {pages} {359} (\bibinfo {year} {2011})},\ \Eprint
  {http://arxiv.org/abs/1012.3049} {arXiv:1012.3049 [nucl-th]} \BibitemShut
  {NoStop}%
\bibitem [{\citenamefont {Boeglin}\ \emph {et~al.}(2011)\citenamefont {Boeglin}
  \emph {et~al.}}]{Boeglin:2011mt}%
  \BibitemOpen
  \bibfield  {author} {\bibinfo {author} {\bibfnamefont {W.~U.}\ \bibnamefont
  {Boeglin}} \emph {et~al.} (\bibinfo {collaboration} {Hall A}),\ }\href
  {\doibase 10.1103/PhysRevLett.107.262501} {\bibfield  {journal} {\bibinfo
  {journal} {Phys. Rev. Lett.}\ }\textbf {\bibinfo {volume} {107}},\ \bibinfo
  {pages} {262501} (\bibinfo {year} {2011})},\ \Eprint
  {http://arxiv.org/abs/1106.0275} {arXiv:1106.0275 [nucl-ex]} \BibitemShut
  {NoStop}%
\bibitem [{\citenamefont {Sargsian}(2001)}]{Sargsian:2001ax}%
  \BibitemOpen
  \bibfield  {author} {\bibinfo {author} {\bibfnamefont {M.~M.}\ \bibnamefont
  {Sargsian}},\ }\href {\doibase 10.1142/S0218301301000617} {\bibfield
  {journal} {\bibinfo  {journal} {Int. J. Mod. Phys.}\ }\textbf {\bibinfo
  {volume} {E10}},\ \bibinfo {pages} {405} (\bibinfo {year} {2001})},\ \Eprint
  {http://arxiv.org/abs/nucl-th/0110053} {arXiv:nucl-th/0110053 [nucl-th]}
  \BibitemShut {NoStop}%
\bibitem [{\citenamefont {Frankfurt}\ \emph {et~al.}(1997)\citenamefont
  {Frankfurt}, \citenamefont {Sargsian},\ and\ \citenamefont
  {Strikman}}]{Frankfurt:1996xx}%
  \BibitemOpen
  \bibfield  {author} {\bibinfo {author} {\bibfnamefont {L.~L.}\ \bibnamefont
  {Frankfurt}}, \bibinfo {author} {\bibfnamefont {M.~M.}\ \bibnamefont
  {Sargsian}}, \ and\ \bibinfo {author} {\bibfnamefont {M.~I.}\ \bibnamefont
  {Strikman}},\ }\href {\doibase 10.1103/PhysRevC.56.1124} {\bibfield
  {journal} {\bibinfo  {journal} {Phys. Rev. C}\ }\textbf {\bibinfo {volume}
  {56}},\ \bibinfo {pages} {1124} (\bibinfo {year} {1997})},\ \Eprint
  {http://arxiv.org/abs/nucl-th/9603018} {arXiv:nucl-th/9603018 [nucl-th]}
  \BibitemShut {NoStop}%
\bibitem [{\citenamefont {Jeschonnek}\ and\ \citenamefont
  {Van~Orden}(2008)}]{Jeschonnek:2008zg}%
  \BibitemOpen
  \bibfield  {author} {\bibinfo {author} {\bibfnamefont {S.}~\bibnamefont
  {Jeschonnek}}\ and\ \bibinfo {author} {\bibfnamefont {J.~W.}\ \bibnamefont
  {Van~Orden}},\ }\href {\doibase 10.1103/PhysRevC.78.014007} {\bibfield
  {journal} {\bibinfo  {journal} {Phys. Rev. C}\ }\textbf {\bibinfo {volume}
  {78}},\ \bibinfo {pages} {014007} (\bibinfo {year} {2008})},\ \Eprint
  {http://arxiv.org/abs/0805.3115} {arXiv:0805.3115 [nucl-th]} \BibitemShut
  {NoStop}%
\bibitem [{\citenamefont {Laget}(2005)}]{Laget:2004sm}%
  \BibitemOpen
  \bibfield  {author} {\bibinfo {author} {\bibfnamefont {J.~M.}\ \bibnamefont
  {Laget}},\ }\href {\doibase 10.1016/j.physletb.2005.01.046} {\bibfield
  {journal} {\bibinfo  {journal} {Phys. Lett. B}\ }\textbf {\bibinfo {volume}
  {609}},\ \bibinfo {pages} {49} (\bibinfo {year} {2005})},\ \Eprint
  {http://arxiv.org/abs/nucl-th/0407072} {arXiv:nucl-th/0407072 [nucl-th]}
  \BibitemShut {NoStop}%
\bibitem [{\citenamefont {Sargsian}(2010)}]{Sargsian:2009hf}%
  \BibitemOpen
  \bibfield  {author} {\bibinfo {author} {\bibfnamefont {M.~M.}\ \bibnamefont
  {Sargsian}},\ }\href {\doibase 10.1103/PhysRevC.82.014612} {\bibfield
  {journal} {\bibinfo  {journal} {Phys. Rev.}\ }\textbf {\bibinfo {volume}
  {C82}},\ \bibinfo {pages} {014612} (\bibinfo {year} {2010})},\ \Eprint
  {http://arxiv.org/abs/0910.2016} {arXiv:0910.2016 [nucl-th]} \BibitemShut
  {NoStop}%
\bibitem [{\citenamefont {Hen}\ \emph {et~al.}(2014{\natexlab{a}})\citenamefont
  {Hen}, \citenamefont {Weinstein}, \citenamefont {Gilad},\ and\ \citenamefont
  {Boeglin}}]{Hen:2014gna}%
  \BibitemOpen
  \bibfield  {author} {\bibinfo {author} {\bibfnamefont {O.}~\bibnamefont
  {Hen}}, \bibinfo {author} {\bibfnamefont {L.~B.}\ \bibnamefont {Weinstein}},
  \bibinfo {author} {\bibfnamefont {S.}~\bibnamefont {Gilad}}, \ and\ \bibinfo
  {author} {\bibfnamefont {W.}~\bibnamefont {Boeglin}},\ }\href@noop {}
  {\enquote {\bibinfo {title} {{Proton and Neutron Momentum Distributions in A
  = 3 Asymmetric Nuclei}},}\ } (\bibinfo {year} {2014}{\natexlab{a}}),\ \Eprint
  {http://arxiv.org/abs/1410.4451} {arXiv:1410.4451 [nucl-ex]} \BibitemShut
  {NoStop}%
\bibitem [{\citenamefont {Sargsian}\ \emph {et~al.}(2003)\citenamefont
  {Sargsian} \emph {et~al.}}]{Sargsian:2002wc}%
  \BibitemOpen
  \bibfield  {author} {\bibinfo {author} {\bibfnamefont {M.~M.}\ \bibnamefont
  {Sargsian}} \emph {et~al.},\ }\href {\doibase 10.1088/0954-3899/29/3/201}
  {\bibfield  {journal} {\bibinfo  {journal} {J. Phys.}\ }\textbf {\bibinfo
  {volume} {G29}},\ \bibinfo {pages} {R1} (\bibinfo {year} {2003})},\ \Eprint
  {http://arxiv.org/abs/nucl-th/0210025} {arXiv:nucl-th/0210025 [nucl-th]}
  \BibitemShut {NoStop}%
\bibitem [{\citenamefont {Cruz-Torres}\ \emph {et~al.}(2019)\citenamefont
  {Cruz-Torres} \emph {et~al.}}]{Cruz-Torres:2019bqw}%
  \BibitemOpen
  \bibfield  {author} {\bibinfo {author} {\bibfnamefont {R.}~\bibnamefont
  {Cruz-Torres}} \emph {et~al.} (\bibinfo {collaboration} {Jefferson Lab Hall A
  Tritium}),\ }\href {\doibase 10.1016/j.physletb.2019.134890} {\bibfield
  {journal} {\bibinfo  {journal} {Phys. Lett.}\ }\textbf {\bibinfo {volume}
  {B797}},\ \bibinfo {pages} {134890} (\bibinfo {year} {2019})},\ \Eprint
  {http://arxiv.org/abs/1902.06358} {arXiv:1902.06358 [nucl-ex]} \BibitemShut
  {NoStop}%
\bibitem [{\citenamefont {Lomon}(2006)}]{Lomon:2006xb}%
  \BibitemOpen
  \bibfield  {author} {\bibinfo {author} {\bibfnamefont {E.~L.}\ \bibnamefont
  {Lomon}},\ }\href@noop {} {\  (\bibinfo {year} {2006})},\ \Eprint
  {http://arxiv.org/abs/nucl-th/0609020} {arXiv:nucl-th/0609020 [nucl-th]}
  \BibitemShut {NoStop}%
\bibitem [{Sim()}]{Simc}%
  \BibitemOpen
  \href@noop {} {\enquote {\bibinfo {title} {{SIMC}},}\ }\bibinfo
  {howpublished}
  {{\url{https://hallcweb.jlab.org/wiki/index.php/SIMC_Monte_Carlo}}},\
  \bibinfo {note} {{Accessed: 2018-10-11}}\BibitemShut {NoStop}%
\bibitem [{\citenamefont {Carasco}\ \emph {et~al.}(2003)\citenamefont {Carasco}
  \emph {et~al.}}]{CARASCO200341}%
  \BibitemOpen
  \bibfield  {author} {\bibinfo {author} {\bibfnamefont {C.}~\bibnamefont
  {Carasco}} \emph {et~al.},\ }\href {\doibase
  https://doi.org/10.1016/S0370-2693(03)00306-X} {\bibfield  {journal}
  {\bibinfo  {journal} {Physics Letters B}\ }\textbf {\bibinfo {volume}
  {559}},\ \bibinfo {pages} {41 } (\bibinfo {year} {2003})}\BibitemShut
  {NoStop}%
\bibitem [{\citenamefont {Bermuth}\ \emph {et~al.}(2003)\citenamefont {Bermuth}
  \emph {et~al.}}]{BERMUTH2003199}%
  \BibitemOpen
  \bibfield  {author} {\bibinfo {author} {\bibfnamefont {J.}~\bibnamefont
  {Bermuth}} \emph {et~al.},\ }\href {\doibase
  https://doi.org/10.1016/S0370-2693(03)00725-1} {\bibfield  {journal}
  {\bibinfo  {journal} {Physics Letters B}\ }\textbf {\bibinfo {volume}
  {564}},\ \bibinfo {pages} {199 } (\bibinfo {year} {2003})}\BibitemShut
  {NoStop}%
\bibitem [{\citenamefont {Ciofi~degli Atti}\ and\ \citenamefont
  {Kaptari}(2005{\natexlab{a}})}]{CiofidegliAtti:2004jg}%
  \BibitemOpen
  \bibfield  {author} {\bibinfo {author} {\bibfnamefont {C.}~\bibnamefont
  {Ciofi~degli Atti}}\ and\ \bibinfo {author} {\bibfnamefont {L.~P.}\
  \bibnamefont {Kaptari}},\ }\href {\doibase 10.1103/PhysRevC.71.024005}
  {\bibfield  {journal} {\bibinfo  {journal} {Phys. Rev. C}\ }\textbf {\bibinfo
  {volume} {71}},\ \bibinfo {pages} {024005} (\bibinfo {year}
  {2005}{\natexlab{a}})},\ \Eprint {http://arxiv.org/abs/nucl-th/0407024}
  {arXiv:nucl-th/0407024 [nucl-th]} \BibitemShut {NoStop}%
\bibitem [{\citenamefont {De~Forest}(1983)}]{DeForest:1983ahx}%
  \BibitemOpen
  \bibfield  {author} {\bibinfo {author} {\bibfnamefont {T.}~\bibnamefont
  {De~Forest}},\ }\href {\doibase 10.1016/0375-9474(83)90124-0} {\bibfield
  {journal} {\bibinfo  {journal} {Nucl. Phys. A}\ }\textbf {\bibinfo {volume}
  {392}},\ \bibinfo {pages} {232} (\bibinfo {year} {1983})}\BibitemShut
  {NoStop}%
\bibitem [{\citenamefont {Machleidt}(2001)}]{Machleidt:2000ge}%
  \BibitemOpen
  \bibfield  {author} {\bibinfo {author} {\bibfnamefont {R.}~\bibnamefont
  {Machleidt}},\ }\href {\doibase 10.1103/PhysRevC.63.024001} {\bibfield
  {journal} {\bibinfo  {journal} {Phys. Rev.}\ }\textbf {\bibinfo {volume}
  {C63}},\ \bibinfo {pages} {024001} (\bibinfo {year} {2001})},\ \Eprint
  {http://arxiv.org/abs/nucl-th/0006014} {arXiv:nucl-th/0006014 [nucl-th]}
  \BibitemShut {NoStop}%
\bibitem [{\citenamefont {Wiringa}\ \emph {et~al.}(1995)\citenamefont
  {Wiringa}, \citenamefont {Stoks},\ and\ \citenamefont
  {Schiavilla}}]{Wiringa:1994wb}%
  \BibitemOpen
  \bibfield  {author} {\bibinfo {author} {\bibfnamefont {R.~B.}\ \bibnamefont
  {Wiringa}}, \bibinfo {author} {\bibfnamefont {V.~G.~J.}\ \bibnamefont
  {Stoks}}, \ and\ \bibinfo {author} {\bibfnamefont {R.}~\bibnamefont
  {Schiavilla}},\ }\href {\doibase 10.1103/PhysRevC.51.38} {\bibfield
  {journal} {\bibinfo  {journal} {Phys. Rev. C}\ }\textbf {\bibinfo {volume}
  {51}},\ \bibinfo {pages} {38} (\bibinfo {year} {1995})},\ \Eprint
  {http://arxiv.org/abs/nucl-th/9408016} {arXiv:nucl-th/9408016 [nucl-th]}
  \BibitemShut {NoStop}%
\bibitem [{\citenamefont {Marcucci}\ \emph {et~al.}(2019)\citenamefont
  {Marcucci}, \citenamefont {Sammarruca}, \citenamefont {Viviani},\ and\
  \citenamefont {Machleidt}}]{Marcucci:2018llz}%
  \BibitemOpen
  \bibfield  {author} {\bibinfo {author} {\bibfnamefont {L.~E.}\ \bibnamefont
  {Marcucci}}, \bibinfo {author} {\bibfnamefont {F.}~\bibnamefont
  {Sammarruca}}, \bibinfo {author} {\bibfnamefont {M.}~\bibnamefont {Viviani}},
  \ and\ \bibinfo {author} {\bibfnamefont {R.}~\bibnamefont {Machleidt}},\
  }\href {\doibase 10.1103/PhysRevC.99.034003} {\bibfield  {journal} {\bibinfo
  {journal} {Phys. Rev.}\ }\textbf {\bibinfo {volume} {C99}},\ \bibinfo {pages}
  {034003} (\bibinfo {year} {2019})},\ \Eprint
  {http://arxiv.org/abs/1809.01849} {arXiv:1809.01849 [nucl-th]} \BibitemShut
  {NoStop}%
\bibitem [{\citenamefont {Ciofi~degli Atti}\ and\ \citenamefont
  {Kaptari}(2005{\natexlab{b}})}]{CiofidegliAtti:2005qt}%
  \BibitemOpen
  \bibfield  {author} {\bibinfo {author} {\bibfnamefont {C.}~\bibnamefont
  {Ciofi~degli Atti}}\ and\ \bibinfo {author} {\bibfnamefont {L.~P.}\
  \bibnamefont {Kaptari}},\ }\href {\doibase 10.1103/PhysRevLett.95.052502}
  {\bibfield  {journal} {\bibinfo  {journal} {Phys. Rev. Lett.}\ }\textbf
  {\bibinfo {volume} {95}},\ \bibinfo {pages} {052502} (\bibinfo {year}
  {2005}{\natexlab{b}})},\ \Eprint {http://arxiv.org/abs/nucl-th/0502045}
  {arXiv:nucl-th/0502045 [nucl-th]} \BibitemShut {NoStop}%
\bibitem [{\citenamefont {Frankfurt}\ \emph {et~al.}(2008)\citenamefont
  {Frankfurt}, \citenamefont {Sargsian},\ and\ \citenamefont
  {Strikman}}]{Frankfurt:2008zv}%
  \BibitemOpen
  \bibfield  {author} {\bibinfo {author} {\bibfnamefont {L.}~\bibnamefont
  {Frankfurt}}, \bibinfo {author} {\bibfnamefont {M.}~\bibnamefont {Sargsian}},
  \ and\ \bibinfo {author} {\bibfnamefont {M.}~\bibnamefont {Strikman}},\
  }\href {\doibase 10.1142/S0217751X08041207} {\bibfield  {journal} {\bibinfo
  {journal} {Int. J. Mod. Phys. A}\ }\textbf {\bibinfo {volume} {23}},\
  \bibinfo {pages} {2991} (\bibinfo {year} {2008})},\ \Eprint
  {http://arxiv.org/abs/0806.4412} {arXiv:0806.4412 [nucl-th]} \BibitemShut
  {NoStop}%
\bibitem [{\citenamefont {Alvioli}\ \emph {et~al.}(2010)\citenamefont
  {Alvioli}, \citenamefont {Ciofi~degli Atti},\ and\ \citenamefont
  {Kaptari}}]{Alvioli:2009zy}%
  \BibitemOpen
  \bibfield  {author} {\bibinfo {author} {\bibfnamefont {M.}~\bibnamefont
  {Alvioli}}, \bibinfo {author} {\bibfnamefont {C.}~\bibnamefont {Ciofi~degli
  Atti}}, \ and\ \bibinfo {author} {\bibfnamefont {L.~P.}\ \bibnamefont
  {Kaptari}},\ }\href {\doibase 10.1103/PhysRevC.81.021001} {\bibfield
  {journal} {\bibinfo  {journal} {Phys. Rev.}\ }\textbf {\bibinfo {volume}
  {C81}},\ \bibinfo {pages} {021001} (\bibinfo {year} {2010})},\ \Eprint
  {http://arxiv.org/abs/0904.4045} {arXiv:0904.4045 [nucl-th]} \BibitemShut
  {NoStop}%
\bibitem [{Sar()}]{SargsPRivate}%
  \BibitemOpen
  \href@noop {} {\enquote {\bibinfo {title} {{M. Sargsian, Private
  communication}},}\ }\BibitemShut {NoStop}%
\bibitem [{\citenamefont {Sargsian}\ \emph
  {et~al.}(2005{\natexlab{a}})\citenamefont {Sargsian}, \citenamefont
  {Abrahamyan}, \citenamefont {Strikman},\ and\ \citenamefont
  {Frankfurt}}]{misak05a}%
  \BibitemOpen
  \bibfield  {author} {\bibinfo {author} {\bibfnamefont {M.~M.}\ \bibnamefont
  {Sargsian}}, \bibinfo {author} {\bibfnamefont {T.~V.}\ \bibnamefont
  {Abrahamyan}}, \bibinfo {author} {\bibfnamefont {M.~I.}\ \bibnamefont
  {Strikman}}, \ and\ \bibinfo {author} {\bibfnamefont {L.~L.}\ \bibnamefont
  {Frankfurt}},\ }\href@noop {} {\bibfield  {journal} {\bibinfo  {journal}
  {Phys. Rev. C}\ }\textbf {\bibinfo {volume} {71}},\ \bibinfo {pages} {044614}
  (\bibinfo {year} {2005}{\natexlab{a}})}\BibitemShut {NoStop}%
\bibitem [{\citenamefont {Sargsian}\ \emph
  {et~al.}(2005{\natexlab{b}})\citenamefont {Sargsian}, \citenamefont
  {Abrahamyan}, \citenamefont {Strikman},\ and\ \citenamefont
  {Frankfurt}}]{misak05b}%
  \BibitemOpen
  \bibfield  {author} {\bibinfo {author} {\bibfnamefont {M.~M.}\ \bibnamefont
  {Sargsian}}, \bibinfo {author} {\bibfnamefont {T.~V.}\ \bibnamefont
  {Abrahamyan}}, \bibinfo {author} {\bibfnamefont {M.~I.}\ \bibnamefont
  {Strikman}}, \ and\ \bibinfo {author} {\bibfnamefont {L.~L.}\ \bibnamefont
  {Frankfurt}},\ }\href@noop {} {\bibfield  {journal} {\bibinfo  {journal}
  {Phys. Rev. C}\ }\textbf {\bibinfo {volume} {71}},\ \bibinfo {pages} {044615}
  (\bibinfo {year} {2005}{\natexlab{b}})}\BibitemShut {NoStop}%
\bibitem [{\citenamefont {Wiringa}\ \emph {et~al.}(2014)\citenamefont
  {Wiringa}, \citenamefont {Schiavilla}, \citenamefont {Pieper},\ and\
  \citenamefont {Carlson}}]{Wiringa:2014}%
  \BibitemOpen
  \bibfield  {author} {\bibinfo {author} {\bibfnamefont {R.~B.}\ \bibnamefont
  {Wiringa}}, \bibinfo {author} {\bibfnamefont {R.}~\bibnamefont {Schiavilla}},
  \bibinfo {author} {\bibfnamefont {S.~C.}\ \bibnamefont {Pieper}}, \ and\
  \bibinfo {author} {\bibfnamefont {J.}~\bibnamefont {Carlson}},\ }\href
  {\doibase 10.1103/PhysRevC.89.024305} {\bibfield  {journal} {\bibinfo
  {journal} {Phys. Rev. C}\ }\textbf {\bibinfo {volume} {89}},\ \bibinfo
  {pages} {024305} (\bibinfo {year} {2014})}\BibitemShut {NoStop}%
\bibitem [{\citenamefont {Hen}\ \emph {et~al.}(2017)\citenamefont {Hen},
  \citenamefont {Miller}, \citenamefont {Piasetzky},\ and\ \citenamefont
  {Weinstein}}]{Hen:2016kwk}%
  \BibitemOpen
  \bibfield  {author} {\bibinfo {author} {\bibfnamefont {O.}~\bibnamefont
  {Hen}}, \bibinfo {author} {\bibfnamefont {G.~A.}\ \bibnamefont {Miller}},
  \bibinfo {author} {\bibfnamefont {E.}~\bibnamefont {Piasetzky}}, \ and\
  \bibinfo {author} {\bibfnamefont {L.~B.}\ \bibnamefont {Weinstein}},\ }\href
  {\doibase 10.1103/RevModPhys.89.045002} {\bibfield  {journal} {\bibinfo
  {journal} {Rev. Mod. Phys.}\ }\textbf {\bibinfo {volume} {89}},\ \bibinfo
  {pages} {045002} (\bibinfo {year} {2017})},\ \Eprint
  {http://arxiv.org/abs/1611.09748} {arXiv:1611.09748 [nucl-ex]} \BibitemShut
  {NoStop}%
\bibitem [{\citenamefont {Ciofi~degli Atti}(2015)}]{Atti:2015eda}%
  \BibitemOpen
  \bibfield  {author} {\bibinfo {author} {\bibfnamefont {C.}~\bibnamefont
  {Ciofi~degli Atti}},\ }\href {\doibase 10.1016/j.physrep.2015.06.002}
  {\bibfield  {journal} {\bibinfo  {journal} {Phys. Rept.}\ }\textbf {\bibinfo
  {volume} {590}},\ \bibinfo {pages} {1} (\bibinfo {year} {2015})}\BibitemShut
  {NoStop}%
\bibitem [{\citenamefont {Piasetzky}\ \emph {et~al.}(2006)\citenamefont
  {Piasetzky}, \citenamefont {Sargsian}, \citenamefont {Frankfurt},
  \citenamefont {Strikman},\ and\ \citenamefont {Watson}}]{piasetzky06}%
  \BibitemOpen
  \bibfield  {author} {\bibinfo {author} {\bibfnamefont {E.}~\bibnamefont
  {Piasetzky}}, \bibinfo {author} {\bibfnamefont {M.}~\bibnamefont {Sargsian}},
  \bibinfo {author} {\bibfnamefont {L.}~\bibnamefont {Frankfurt}}, \bibinfo
  {author} {\bibfnamefont {M.}~\bibnamefont {Strikman}}, \ and\ \bibinfo
  {author} {\bibfnamefont {J.~W.}\ \bibnamefont {Watson}},\ }\href {\doibase
  10.1103/PhysRevLett.97.162504} {\bibfield  {journal} {\bibinfo  {journal}
  {Phys. Rev. Lett.}\ }\textbf {\bibinfo {volume} {97}},\ \bibinfo {pages}
  {162504} (\bibinfo {year} {2006})}\BibitemShut {NoStop}%
\bibitem [{\citenamefont {Subedi}\ \emph {et~al.}(2008)\citenamefont {Subedi}
  \emph {et~al.}}]{subedi08}%
  \BibitemOpen
  \bibfield  {author} {\bibinfo {author} {\bibfnamefont {R.}~\bibnamefont
  {Subedi}} \emph {et~al.},\ }\href@noop {} {\bibfield  {journal} {\bibinfo
  {journal} {Science}\ }\textbf {\bibinfo {volume} {320}},\ \bibinfo {pages}
  {1476} (\bibinfo {year} {2008})}\BibitemShut {NoStop}%
\bibitem [{\citenamefont {Korover}\ \emph {et~al.}(2014)\citenamefont
  {Korover}, \citenamefont {Muangma}, \citenamefont {Hen} \emph
  {et~al.}}]{korover14}%
  \BibitemOpen
  \bibfield  {author} {\bibinfo {author} {\bibfnamefont {I.}~\bibnamefont
  {Korover}}, \bibinfo {author} {\bibfnamefont {N.}~\bibnamefont {Muangma}},
  \bibinfo {author} {\bibfnamefont {O.}~\bibnamefont {Hen}},  \emph {et~al.},\
  }\href {\doibase 10.1103/PhysRevLett.113.022501} {\bibfield  {journal}
  {\bibinfo  {journal} {Phys. Rev. Lett.}\ }\textbf {\bibinfo {volume} {113}},\
  \bibinfo {pages} {022501} (\bibinfo {year} {2014})}\BibitemShut {NoStop}%
\bibitem [{\citenamefont {Hen}\ \emph {et~al.}(2014{\natexlab{b}})\citenamefont
  {Hen} \emph {et~al.}}]{Hen:2014nza}%
  \BibitemOpen
  \bibfield  {author} {\bibinfo {author} {\bibfnamefont {O.}~\bibnamefont
  {Hen}} \emph {et~al.},\ }\href {\doibase 10.1126/science.1256785} {\bibfield
  {journal} {\bibinfo  {journal} {Science}\ }\textbf {\bibinfo {volume}
  {346}},\ \bibinfo {pages} {614} (\bibinfo {year} {2014}{\natexlab{b}})},\
  \Eprint {http://arxiv.org/abs/1412.0138} {arXiv:1412.0138 [nucl-ex]}
  \BibitemShut {NoStop}%
\bibitem [{\citenamefont {Cohen}\ \emph {et~al.}(2018)\citenamefont {Cohen}
  \emph {et~al.}}]{Cohen:2018gzh}%
  \BibitemOpen
  \bibfield  {author} {\bibinfo {author} {\bibfnamefont {E.~O.}\ \bibnamefont
  {Cohen}} \emph {et~al.} (\bibinfo {collaboration} {CLAS}),\ }\href {\doibase
  10.1103/PhysRevLett.121.092501} {\bibfield  {journal} {\bibinfo  {journal}
  {Phys. Rev. Lett.}\ }\textbf {\bibinfo {volume} {121}},\ \bibinfo {pages}
  {092501} (\bibinfo {year} {2018})},\ \Eprint
  {http://arxiv.org/abs/1805.01981} {arXiv:1805.01981 [nucl-ex]} \BibitemShut
  {NoStop}%
\bibitem [{\citenamefont {Duer}\ \emph {et~al.}(2018)\citenamefont {Duer} \emph
  {et~al.}}]{Duer:2018sby}%
  \BibitemOpen
  \bibfield  {author} {\bibinfo {author} {\bibfnamefont {M.}~\bibnamefont
  {Duer}} \emph {et~al.} (\bibinfo {collaboration} {CLAS}),\ }\href {\doibase
  10.1038/s41586-018-0400-z} {\bibfield  {journal} {\bibinfo  {journal}
  {Nature}\ }\textbf {\bibinfo {volume} {560}},\ \bibinfo {pages} {617}
  (\bibinfo {year} {2018})}\BibitemShut {NoStop}%
\bibitem [{\citenamefont {Duer}\ \emph {et~al.}(2019)\citenamefont {Duer} \emph
  {et~al.}}]{Duer:2018sxh}%
  \BibitemOpen
  \bibfield  {author} {\bibinfo {author} {\bibfnamefont {M.}~\bibnamefont
  {Duer}} \emph {et~al.} (\bibinfo {collaboration} {CLAS Collaboration}),\
  }\href {\doibase 10.1103/PhysRevLett.122.172502} {\bibfield  {journal}
  {\bibinfo  {journal} {Phys. Rev. Lett.}\ }\textbf {\bibinfo {volume} {122}},\
  \bibinfo {pages} {172502} (\bibinfo {year} {2019})}\BibitemShut {NoStop}%
\bibitem [{\citenamefont {Pybus}\ \emph {et~al.}(2020)\citenamefont {Pybus},
  \citenamefont {Korover}, \citenamefont {Weiss}, \citenamefont {Schmidt},
  \citenamefont {Barnea}, \citenamefont {Higinbotham}, \citenamefont
  {Piasetzky}, \citenamefont {Strikman}, \citenamefont {Weinstein},\ and\
  \citenamefont {Hen}}]{Pybus:2020itv}%
  \BibitemOpen
  \bibfield  {author} {\bibinfo {author} {\bibfnamefont {J.~R.}\ \bibnamefont
  {Pybus}}, \bibinfo {author} {\bibfnamefont {I.}~\bibnamefont {Korover}},
  \bibinfo {author} {\bibfnamefont {R.}~\bibnamefont {Weiss}}, \bibinfo
  {author} {\bibfnamefont {A.}~\bibnamefont {Schmidt}}, \bibinfo {author}
  {\bibfnamefont {N.}~\bibnamefont {Barnea}}, \bibinfo {author} {\bibfnamefont
  {D.~W.}\ \bibnamefont {Higinbotham}}, \bibinfo {author} {\bibfnamefont
  {E.}~\bibnamefont {Piasetzky}}, \bibinfo {author} {\bibfnamefont
  {M.}~\bibnamefont {Strikman}}, \bibinfo {author} {\bibfnamefont {L.~B.}\
  \bibnamefont {Weinstein}}, \ and\ \bibinfo {author} {\bibfnamefont
  {O.}~\bibnamefont {Hen}},\ }\href@noop {} {\  (\bibinfo {year} {2020})},\
  \Eprint {http://arxiv.org/abs/2003.02318} {arXiv:2003.02318 [nucl-th]}
  \BibitemShut {NoStop}%
\bibitem [{\citenamefont {Schmidt}\ \emph {et~al.}(2020)\citenamefont {Schmidt}
  \emph {et~al.}}]{schmidt20}%
  \BibitemOpen
  \bibfield  {author} {\bibinfo {author} {\bibfnamefont {A.}~\bibnamefont
  {Schmidt}} \emph {et~al.} (\bibinfo {collaboration} {CLAS Collaboration}),\
  }\href@noop {} {\bibfield  {journal} {\bibinfo  {journal} {Nature}\ }\textbf
  {\bibinfo {volume} {578}},\ \bibinfo {pages} {540–544} (\bibinfo {year}
  {2020})}\BibitemShut {NoStop}%
\end{thebibliography}%

\end{document}